\documentclass{jfm}
\usepackage{natbib}
\usepackage{graphicx}
\usepackage{bm}

\usepackage{xcolor}

\title{Information transfer between turbulent boundary layer and porous media}
\author{Wenkang Wang\aff{1,2}, Xu Chu\aff{1,3}
  \corresp{\email{xu.chu@itlr.uni-stuttgart.de}}, Adri\'{a}n Lozano-Dur\'{a}n\aff{4},
  Rainer Helmig\aff{2} \and Bernhard Weigand \aff{1}}

\affiliation{
\aff{1} Institute of Aerospace Thermodynamics, University of Stuttgart, Germany
\aff{2} Institute for Modelling Hydraulic and Environmental Systems , University of Stuttgart, Germany
\aff{3} Stuttgart Center of Simulation Science (SimTech), University of Stuttgart, Germany
\aff{4} Center for Turbulence Research, Stanford University, Stanford, CA 94305, USA
}

\begin{document}

\maketitle
\begin{abstract}
    The interaction between the flow above and below a permeable wall
    is a central topic in the study of porous media. While previous
    investigations have provided compelling evidence of the strong
    coupling between the two regions, few studies have quantitatively
    measured the directionality, i.e., causal-and-effect relations, of
    this interaction. To shed light on the problem, we use transfer
    entropy as a marker to evaluate the causal interaction between the
    free turbulent flow and the porous media using interface-resolved
    direct numerical simulation. Our results show
      that the porosity of the porous medium has a profound impact on
      the intensity, time scale and spatial extent of
      surface-subsurface interactions. For values of porosity equal to
      0.5, top-down and bottom-up interactions strongly are
      asymmetric, the former being mostly influenced by small
      near-wall eddies.  As the porosity increases, both top-down and
      bottom-up interactions are dominated by shear-flow
      instabilities.
\end{abstract}

\section{Introduction}

Turbulent shear flows over porous walls are frequently encountered in
the natural environment and industrial applications, such as oil
wells, catalytic reactors, heat exchangers, and porous river beds, to
name a few.  As the blocking effect of the wall is relaxed, the porous
bed allows for mass, momentum and energy exchange across the permeable
interface, which triggers changes in both the surface and subsurface
flows~\citep[]{breugem2006influence, manes2011turbulent,
  kim2020experimental, fang2018influence, suga_okazaki_kuwata_2020}.
A widely observed form of interaction in flows over porous beds is the
upwelling and downwelling motion across the permeable interface, which
enables the exchange of information between surface and subsurface
flow. Studies relying on correlation analysis, conditional averaging, and modal analysis
have revealed that the transport of wall-normal fluid across the
interface is accompanied by streamwise velocity fluctuations in
surface flow~\citep[]{breugem2006influence, kim2020experimental}. Moreover, the turbulence intensity below the
interface is also modulated by outer large-scale motions
\cite[]{Efstathiou.2018, kim2020experimental}.  The influence of the
porous bed is not constrained to the near-wall region. It has been
observed that large-scale vortical structures emerge in the surface
flow as the wall permeability increases, which has been attributed to
Kelvin-Helmholtz (KH) type of instabilities from inflection points of
the mean velocity profile \citep[]{Jimenez.2001, breugem2006influence,
  manes2011turbulent}. In addition, the streaks and streamwise
vortices characteristic of the near-wall cycle are substantially
weakened in the presence of porous walls compared to smooth-wall
turbulence.  This suggest that the flow structure over porous walls is
is governed by the balance of two competing mechanism: the formation
of wall-attached eddies and the disruption by shear layer
instabilities~\cite[]{manes2011turbulent, Efstathiou.2018}.

Despite the last advancements in the field, the dynamics of the mass
and momentum transfer across the interface of porous media is far
from being settled. Although we do have a solid knowledge on the flow
structures resulting from the surface-subsurface interaction, the
fundamental cause-effect relations of these fluid motions are rarely
inspected. To what extent does the outer turbulence determines the
flow inside porous media and vice versa?  These questions cannot be
addressed by correlation analysis and traditional mode decomposition
methods (e.g. proper orthogonal decomposition, POD) as these
approaches do not provide the directionality and time asymmetry
required to quantify causation~\citep[]{kantz2004nonlinear,
  pearl2009causality}. Recently, \cite{lozano2020causality}
highlighted the importance of causal inference in fluid mechanics
\citep[see also][]{lozano2020cause} and proposed to leverage
information theoretic metrics to explore causality in turbulent
flows. In particular, \cite{lozano2020causality} examined the causal
interactions of energy-eddies in wall-bounded turbulence using
transfer entropy~\citep{Schreiber2000}. In this study, we conduct
direct numerical simulation (DNS) of a fully-resolved interface and
porous domain and follow the analysis of \cite{lozano2020causality} to
quantify the causality among the interaction of surface and subsurface
flow.

The work is organised as follows: we present the details of the DNS
dataset and the definition of transfer entropy in
\S\ref{sec:numerical}. In \S\ref{sec:results}, we discuss the typical
upwelling/downwelling process and showed that the porosity of the
media has a major impact on the inner and outer flow structures as
well as in their causal relations. Finally, conclusions are offered in
\S\ref{sec:conclusion}.


\section{Numerical simulations and methods} \label{sec:numerical}

\subsection{Numerical setup}

The three-dimensional incompressible Navier$-$Stokes equations are
solved in non-dimensional form, 
\begin{equation}
\frac{\partial u_j}{\partial x_j}=0, \quad
\frac{\partial u_i}{\partial t}+\frac{\partial u_i u_j}{\partial x_j}=
-\frac{\partial p}{\partial x_i}+\frac{1}{Re}\frac{\partial^2 u_i} {\partial x_j \partial x_j}+\Pi \delta_{i1},
\end{equation}
where $\Pi$ is a constant pressure gradient in the mean-flow
direction.  The governing equations are non-dimensionalized by normalizing lengths by half of distance between two cylinders $D/2$ (figure \ref{fig:meanu}a), velocities by the averaged bulk velocity $U_b$ of the free flow region ($y=[0,10]$), such that time is non-dimensionalized by $D/(2U_b$).
The spectral/$hp$ element solver Nektar++
\citep{Cantwell.2011} is used to discretize the numerical domain
containing complex geometrical structures. The solver allows for
arbitrary-order spectral/$hp$ discretisations with hybrid shaped
elements.  The time-stepping is performed with a second-order mixed
implicit-explicit (IMEX) scheme. The time step is fixed to $\Delta T/(D/2U_b)=0.005$.
  

Hereafter, the velocity components in the streamwise $x$, wall normal
$y$ and spanwise $z$ directions are denoted as $u$, $v$ and $w$,
respectively.  The domain size ($L_x \times L_y \times L_z$) is $100
\times 20 \times 8\pi$. The lower half $y=[-10,0]$ contains the porous
media and the upper half $y=[0,10]$ is the free channel-flow. The
  porous layer consists of 50 cylinder elements aligned along the
  streamwise direction and arranged in 5 rows in the wall-normal
  direction as illustrated in figure \ref{fig:meanu}(a). The distance
  between two cylinders is fixed at $D=2$.

The no-slip boundary condition is applied to the cylinders, the upper
wall, and lower wall. Periodic boundary conditions are used in
streamwise and spanwise directions. The geometry is discretised using
quadrilateral elements on the $x-y$ plane with local refinement near
the interface. High-order Lagrange polynomials ($P=4-7$)
are used within each element on the $x-y$ plane. The spanwise direction is extended with a Fourier-based spectral method.

Three DNS cases are performed with varying porosity $\varphi$, which is
defined as the ratio of the void volume to the total volume of the
porous structure. The parameters of simulations are listed in table \ref{tab:1}.  The superscripts $(\cdot)^{\mathrm{p}}$ and
$(\cdot)^{\mathrm{t}}$ represent variables of permeable wall and top smooth wall sides, respectively. Variables with superscript $+$ are scaled by friction velocities $u_\tau$ of their respective side and viscosity $\nu$. The Reynolds number of top wall boundary layer is set to be $Re_\tau^{\mathrm t}=\delta^{\mathrm t}u_\tau^{\mathrm
  t}/\nu\approx180$ for all cases ($\delta$ is the distance between the position of maximum streamwise velocity and the wall). 
In this manner, changes in the top wall
boundary layer are minimized. 
  On the top smooth wall side, the streamwise cell size ranges from $6.3\le\Delta
x^{\mathrm{t}+}\le8.5$ and the spanwise cell size is below $\Delta
z^{\mathrm{t}+}=5.6$. On the porous
media side, $\Delta z^{\mathrm{p}+}$ is below 8.4, whereas $\Delta x^{\mathrm{p}+}$ and $\Delta y^{\mathrm{p}+}$ are enhanced by polynomial refinement of local mesh \citep{Cantwell.2011}. 
%
\begin{table}
\centering
\begin{tabular}{cccccccccc}
case&$\varphi$& $U_b$   &$u_\tau^{\mathrm{p}}$&$Re_{\tau}^{\mathrm{p}}$&$\Delta x^{\mathrm{p}+}$/$\Delta y^{\mathrm{p}+}$/$\Delta z^{\mathrm{p}+}$&$u_\tau^{\mathrm{t}}$&$Re_{\tau}^{\mathrm{t}}$&$\Delta x^{\mathrm{t}+}$/$\Delta y^{\mathrm{t}+}$/$\Delta z^{\mathrm{t}+}$\\
C05 & 0.50 & 4.1   & 0.31 & 336 & 2.2/0.38/5.3 & 0.27 & 180 & 6.3/0.43/4.5 \\
C06 & 0.60 & 4.1   & 0.36 & 464 & 3.5/0.50/6.8 & 0.28 & 190 & 5.5/0.47/ 5.4  \\
C07 & 0.70 & 4.0   & 0.38 & 625 & 4.2/0.55/8.4 & 0.23 & 160 & 5.2/0.51/ 5.1 \\
C08 & 0.80 & 3.0   & 0.38 & 793 & 4.3/0.64/8.0 & 0.21 & 170 & 8.5/0.34 /5.6 \\
\end{tabular}
\caption{Simulation parameters. The porosity of the porous medium
  region is $\varphi$. The friction Reynolds numbers for the porous
  and impermeable top walls are $Re_{\tau}^{\mathrm{p}}$ and
  $Re_{\tau}^{\mathrm{t}}$, respectively. $\Delta
  x^{\mathrm{p}+}$/$\Delta y^{\mathrm{p}+}$/$\Delta z^{\mathrm{p}+}$
  and $\Delta x^{\mathrm{t}+}$/$\Delta y^{\mathrm{t}+}$/$\Delta
  z^{\mathrm{t}+}$ are the respective grid spacings in wall units of
  porous wall and top wall, respectively. }
\label{tab:1}
\end{table}

\subsection{Causality among time signals as transfer entropy}

We use transfer entropy \citep{Schreiber2000} as a marker to evaluate
the direction of coupling, i.e., the cause-effect relationship,
between two time-signals representative of some fluid quantities of
interest~\citep{lozano2020causality}. Transfer entropy is an
information-theoretic metric representing the dependence of the present
of $X$ on the past of $Y$. This dependence is measured as the decrease
in uncertainty (or entropy) of the signal $X$ by knowing the past
state of $Y$. Transfer entropy from $X$ and $Y$ is formally
defined as
\begin{equation}
    T_{Y\rightarrow X}(\Delta t)=H(X_t|X_{t-1})-H(X_t|{X_{t-1},Y_{t-\Delta t}}),
    \label{eq:Tij}
\end{equation}
where the subscript $t$ denotes time and $\Delta t$ is the time-lag to
evaluate causality. $H(A|B)$ is the conditional Shannon entropy of the
variable $A$ given $B$,
\begin{equation}
    H(A|B)=E[\mathrm{log}(p(A,B))]-E[\mathrm{log}(p(B))],
\end{equation}
where $p(\cdot)$ is the probability density function, and $E[\cdot]$
denotes the expected value.

In order to quantify the relative strength of causality, we scale the
magnitude of $T_{Y\rightarrow X}$ within the range $[0,1]$ and define the
normalized transfer entropy~\citep{gourevitch2007evaluating}, 
\begin{equation}
    \tilde T_{Y\rightarrow X}=\frac{T_{Y\rightarrow X}-E[T_{Y^s\rightarrow X}]}{H(X_t|X_{t-1})},
    \label{eq:nte}
\end{equation}
where $H(X_t|X_{t-1})$ is the intrinsic uncertainty of $X$ conditioned
on knowing its own history. Thus, (\ref{eq:nte}) represents the
fraction of information in $X$ not explained by its own past that is
explained by the past of $Y$.  The term $E[T_{Y^s\rightarrow X}]$ is
introduced to alleviate the bias due to statistical errors. This is
achieved with the surrogate variable $Y^s$, which is constructed by
randomly permuting $Y$ in time to break any causal links with $X$.

\section{Results} \label{sec:results}

\subsection{Mean velocity and turbulent kinetic energy profiles}

\begin{figure}
\centering
\includegraphics[width=0.8\textwidth]{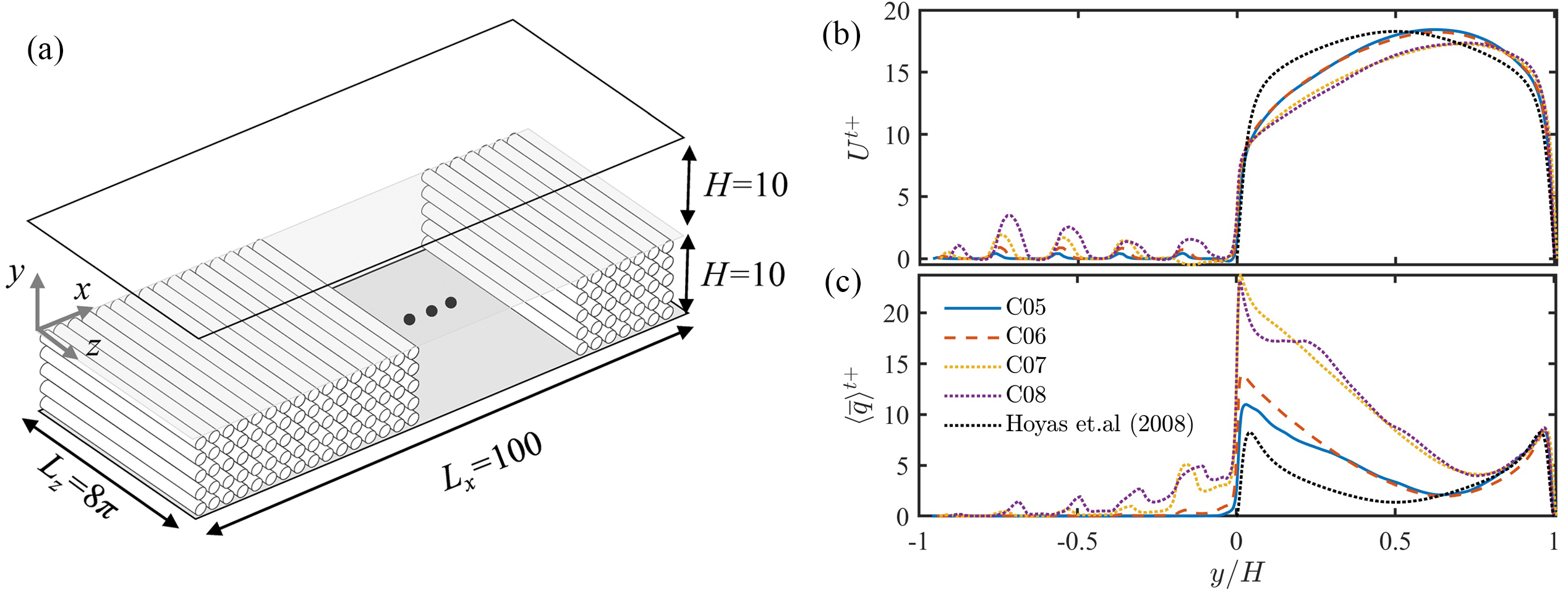}
\caption{(a) Configuration of the computational domain; (b) mean
  streamwise velocity normalized by
  $u_\tau^{\mathrm{t}}$; (c) mean TKE profiles normalized
  by $u_\tau^{\mathrm{t}2}$. Profiles of impermeable smooth wall channel (\cite{hoyas2008reynolds}, $Re_\tau=180$) are superimposed in (b,c) for comparison.}
\label{fig:meanu}
\end{figure}

The mean statistics of the three cases are briefly discussed here to
outline the impact of porosity on surface and subsurface flow. Figure
\ref{fig:meanu}(b,c) shows the mean profiles for the streamwise velocity
$\langle \overline u \rangle$ and turbulent kinetic energy (TKE)
$\langle \overline q\rangle=\langle \overline {u'u'+
  v'v'+w'w'}\rangle/2$ normalized by $u_\tau^t$.
The operators $\langle \cdot\rangle$ and $\overline{(\cdot)}$ represent spatial average in $x$-$z$ and temporal average, respectively, and the
superscript prime denotes turbulent fluctuations
$\phi'=\phi-\overline{\phi}$. The profiles of impermeable smooth wall channel from \cite{hoyas2008reynolds} are also included for comparison. For increasing wall porosity, the mean
velocity profile becomes more skewed toward the top wall as a
consequence of the higher skin friction on the porous
wall~\cite[]{breugem2006influence}. Below the interface ($y=0$), the
mean velocity profiles exhibit a clear inflection point, which is
typically associated with KH-type instabilities responsible for
additional turbulent structures \citep{manes2011turbulent}. This is
partly substantiated by figure \ref{fig:meanu}(c), which shows that
the TKE below and above the interface is significantly enhanced for
highest porosity considered here (case C08). In the following
sections, we illustrate how the change of flow structures in the
interface region affects the interaction between surface and
sub-surface flow.

\subsection{Upwelling and downwelling events}

\begin{figure}
\centering
\includegraphics[width=0.8\textwidth]{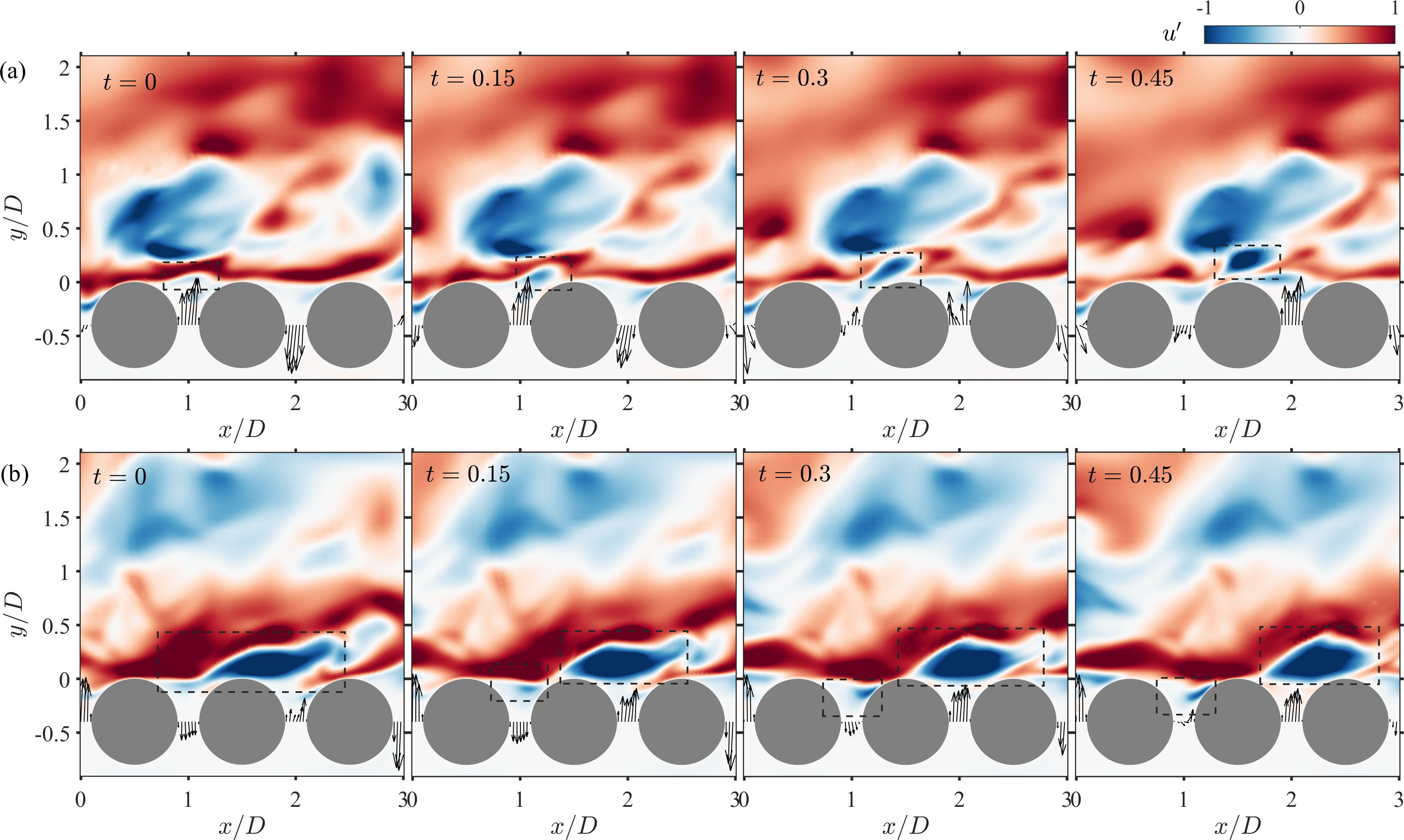}
\caption{Instantaneous $u'$ fields (C05) during (a) an upwelling event
  and (b) a downwelling event. Instantaneous $(u',v')$ at the gap
  between cylinders are illustrated by vectors. }
\label{fig:inst_u}
\end{figure}

The upwelling/downwelling events transport fluid directly across the
interface, which provide an intuitive picture of the interaction
process.
Examples of typical upwelling and downwelling events are captured in
figure \ref{fig:inst_u} for case C05.  During the upwelling event
(figure \ref{fig:inst_u}a, $x/D=1$), a small amount of fluid is
ejected from beneath the interface ($t=0$), which later becomes a low
speed bulb above the interface ($t=0.15$-$0.3$) to finally merged with
a large low-speed structure downstream ($t=0.45$). During the
downwelling event (figure \ref{fig:inst_u}b, $x/D=1$), a portion of
low-speed fluid is absorbed into the porous medium ($t=0$). The
low-speed structure is later separated into two structures as the
upper part is convected downstream ($t=0.15$-$0.45$). The observations
from the instantaneous fields also suggest that upwelling/downwelling
events are subjected to the modulation of large-scale motions in free
flow region. Specifically, the upwelling events at the gaps are often
related to large-scale low-speed structures above the interface (see
figure \ref{fig:inst_u}a at $y/D=0.5$), while downwelling events are
usually associated with high speed structures (see figure
\ref{fig:inst_u}b at $y/D=0.5$). This connection was also observed by
\cite[]{kim2018experimental}, and it is further discussed in the
following sections. In addition, the momentum flux between
neighbouring gaps is usually negatively correlated, i.e., upwelling
and downwelling events usually occur side by side probably due to the
constrain imposed by continuity at the interface.

\subsection{Coherent structure of POD modes} \label{sec:pod}

\begin{figure}
\centering
\includegraphics[width=1\textwidth]{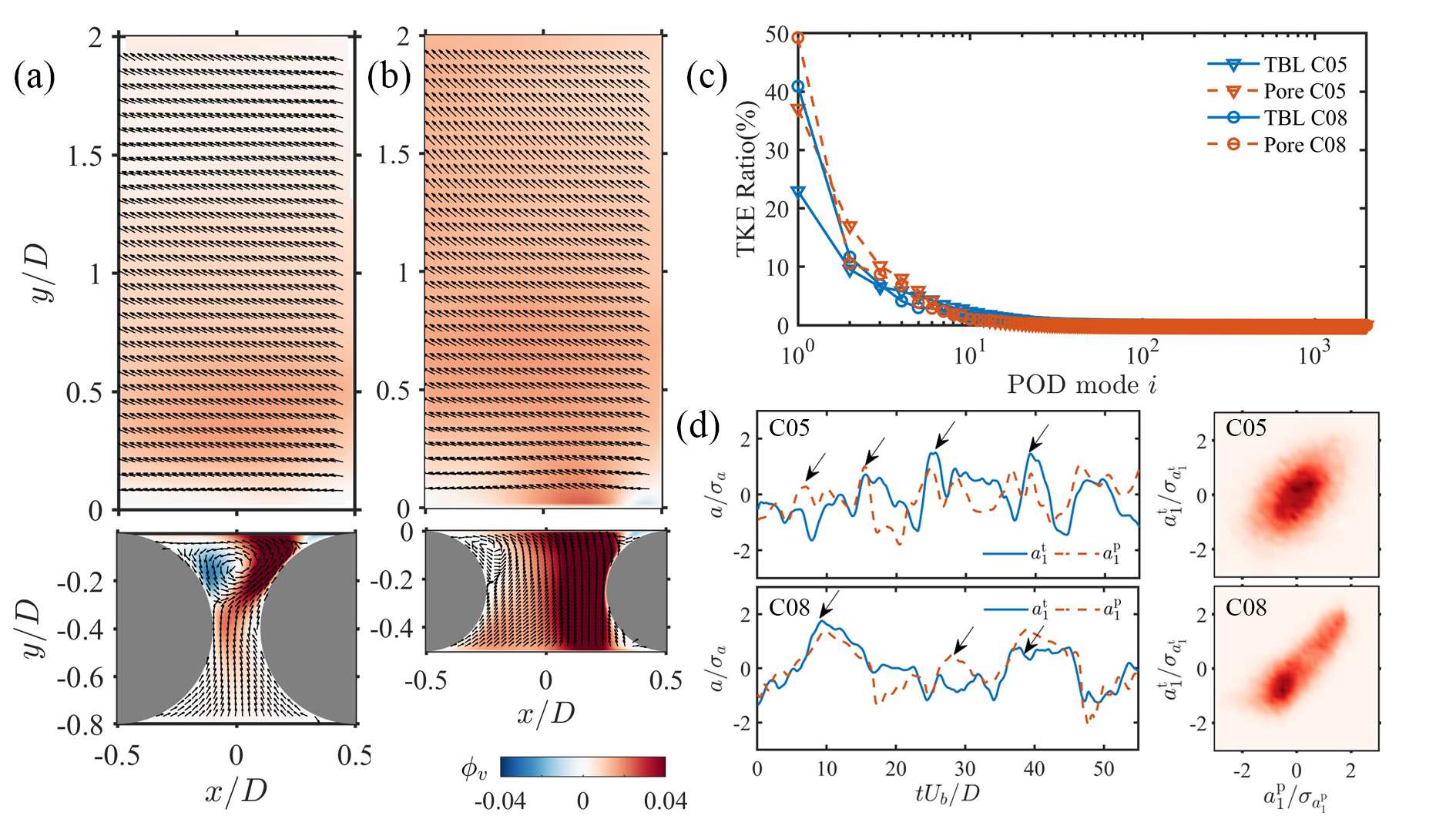}
\caption{First POD spatial mode of TBL $\phi^{\mathrm{t}}_1$ and
  porous region $\phi^{\mathrm{p}}_1$ for cases (a) C05 and (b) C08;
  (c) shows the TKE contribution ratio of mode order $i$;(d) shows the
  time coefficient $a^{\mathrm{t}}_1$ (blue solid line) and
  $a^{\mathrm{p}}_1$ (red dashed line) as well as their joint PDF for
  cases C05 (upper row) and C08 (lower row).}
\label{fig:POD_A}
\end{figure}

To investigate the causal relation between surface and subsurface
flows, we characterize the flow structure within each region using
Proper Orthogonal Decomposition (POD). We extract energetic modes
above and below the interface and use the corresponding time
coefficients to measure causality between interactions. In POD, the
fluctuating velocity field $\bm{u}'(\bm{x}, t)=[u'(\bm{x},
  t),v'(\bm{x}, t),w'(\bm{x}, t)]$, is decomposed via optimal energy
criteria into
 \begin{equation}\label{eq:pod}
 \bm{u}'\left( {\bm{x},t} \right) = \sum\limits_{k = 1}^K {{a_k}\left( t \right){\bm{\phi} _k}\left( \bm{x} \right)}, 
 \end{equation}
where $a_k(t)$ is the time coefficient for the $k^{\mathrm{th}}$ rank
mode and $\bm{\phi}_k(\bm{x})$ are the modes which constitute an
orthogonal spatial basis.  We focus on the local information transfer
in the vicinity of one pore unit. Consistently, the wall-normal extent
of boundary layer region is selected from the crest height ($y=0$) to
the channel center. For the porous medium region, we focus on the 1st
layer of cylinders to capture the so called transition
layer~\citep{kim2020experimental}, i.e., the region with direct mass
and momentum exchange with the free flow.  The first modes for cases C05
and C08, denoted by $\phi^{\mathrm{p}}_1$ and $\phi^{\mathrm{t}}_1$,
respectively (superscript p for `porous' and t for `turbulent boundary
layer' or TBL), are shown in figure \ref{fig:POD_A}.  The POD is
computed from a time-series of of 2000 snapshots spanning a time of
$TU_b/D=150$-$200$. The time interval between snapshot is chosen to be
$\Delta tU_b/D\approx0.1$, which is $\mathcal{O}(10^{-2})$ the time
scale of the typical period of fluctuations inside porous media
(discussed later in figure \ref{fig:POD_A}d). The PODs are computed
for all 50 pore units in streamwise and 12 evenly-spaced positions in
spanwise direction (spanwise interval $\Delta Z^{\mathrm t+}\ge100$). 
The consistency of POD modes at different positions is checked by evaluating the correlation coefficients between the $\phi^{\mathrm{p}}_1$ (or $\phi^{\mathrm{t}}_1$) at different $x$ and $z$ positions, which are at a high level of $0.97\pm0.02$.

For the two porosities shown in figure \ref{fig:POD_A}, the 1st mode
for the TBL region $\phi^{\mathrm{t}}_1$ consists of a large
countergradient region (i.e., Q2 when $u'<0$ and $v'>0$, and Q4 when
$u'<0$ and $v'>0$). Although the streamwise extent of this structure
is unknown due to the limited field of view, the wall-normal extent
reaches the center of the channel and hence it can be deemed as
`large'.
The 1st mode in the porous region $\phi^{\mathrm{p}}$ shows a richer
structure which is dramatically affected by the porosity. In case C05,
$\phi^{\mathrm{p}}_1$ encompasses a vortex trapped on the top of the
cylinder's gap, and a vertical momentum transfer within the gap. In
contrast, the vortex above the gap is considerably smaller in C08,
leaving room for a stronger vertical momentum flux within the gap.
Figure \ref{fig:POD_A}(c) shows that the modes $\phi^{\mathrm{p}}_1$
already account for 40-50\% of the TKE and they are taken as the
simplest representation of the flow patterns close to the interface.

The time coefficients of first modes, $a^{\mathrm{t}}_1$ and
$a^{\mathrm{p}}_1$, are compared in figure \ref{fig:POD_A}(d). Note
that time coefficients represent the instantaneous magnitude of
corresponding spatial mode and, as such, can be used as markers of the
flow structure in each region. In the current context,
$a^{\mathrm{t}}_1$ denotes the intensity of Q2/Q4 events above the
pore unit, whereas $a^{\mathrm{p}}_1$ reflects the strength of the
vortex and momentum flux at the gap. An excerpt of the time series of
$a^{\mathrm{t}}_1$ and $a^{\mathrm{p}}_1$ is plotted in figure
\ref{fig:POD_A}(d)-left.  Overall, both time-signals show a certain
degree of synchronization, which seems to become stronger as porosity
increases. Large-scale fluctuations with period $\Delta TU_b/D\ge10$ can be observed in $a^{\mathrm{t}}_1$ and $a^{\mathrm{p}}_1$ for low porosity case C08, the scale of which is consistent with the prediction of linear stability theory \cite{Jimenez.2001} and experimental observations \cite[]{manes2011turbulent,kim2020experimental}. Figure \ref{fig:POD_A}(d)-right contains the joint
probability density function (JPDF) of $a^{\mathrm{t}}_1$ and
$a^{\mathrm{p}}_1$ and shows that the mode is distributed along
$a^{\mathrm{t}}_1 \sim a^{\mathrm{p}}_1$ for both case C08 and case
C05, with the variables in the former case being clearly more
correlated.  However, the JPDF does not provides information about
the direction of interaction, i.e., whether $a^{\mathrm{t}}_1$ causes
$a^{\mathrm{p}}_1$ or vice versa.

\subsection{Information transfer between free-flow and porous-media structures}\label{sec:TE-POD}
\begin{figure}
\centering
\includegraphics[width=1\textwidth]{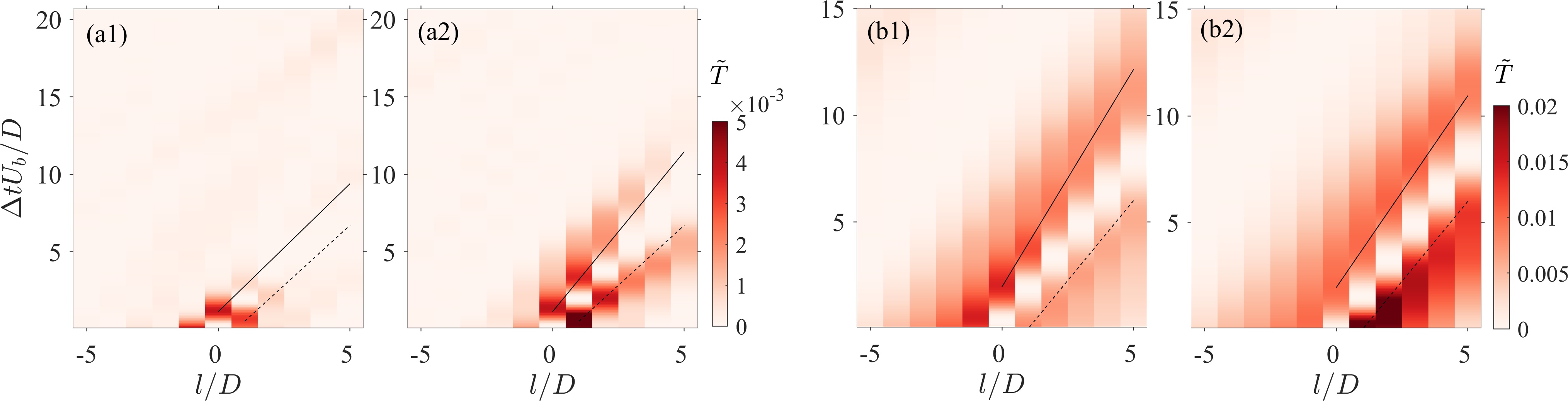}
\caption{Transfer entropy of POD time-coefficients. Case C05: (a1)
  $\tilde T_{a^{\mathrm{t}}_1\rightarrow a^{\mathrm{p}}_1}$; (a2)
  $\tilde T_{a^{\mathrm{p}}_1\rightarrow a^{\mathrm{t}}_1}$. Case C08: (b1)
  $\tilde T_{a^{\mathrm{t}}_1\rightarrow a^{\mathrm{p}}_1}$; (b2)
  $\tilde T_{a^{\mathrm{p}}_1\rightarrow a^{\mathrm{t}}_1}$. Solid and dashed
  lines are used to mark the upper and lower ridges on the contours,
  respectively. }
\label{fig:Tij}
\end{figure}

We assess the causal relation between the coherent structures by
evaluating the transfer entropy between $a^{\mathrm{t}}_1$ and
$a^{\mathrm{p}}_1$. The contour of normalized transfer entropy of the
time coefficients is shown in figure \ref{fig:Tij} as a function of
time-delay $\Delta t$ (i.e., the time-horizon for causality as shown
in figure (\ref{eq:Tij})). The information transfer between different
pore units is also inspected, with $l$
the streamwise offset of target time coefficient
$X=a^{\mathrm{t}}_{1,l}(t)$ (or $a^{\mathrm{p}}_{1,l}(t)$)
from $Y=a^{\mathrm{p}}_1(t)$ (or $a^{\mathrm{t}}_1(t)$).
For example, $l/D=2$ denotes a target
signal $X$ acquired 2 pore units downstream the location of $Y$. We
take into account all 50 pore units in the streamwise direction and 12
spanwise positions, which amounts to a number of samples equal to
$\mathcal{O}(10^6)$ for each case. Adaptive binning method
\citep{darbellay1999estimation} is used to estimate the probability
density function in (\ref{eq:Tij}) with the bin number equal to 50 in
each dimension according to the empirical recommendation of
\cite{hacine2012low}. It was tested that doubling and halving the
number of bins did not alter the conclusions presented hereafter.

The strongest values of $\tilde T$ are organized along the two ridges
in the space of $(l, \Delta t)$ indicated by solid and dashed
lines in figure \ref{fig:Tij}. These ridges mark the time delay and
spatial offset at which information transfer between the two signals
is maximum.  When $X$ and $Y$ are within the same pore unit ($l=0$), the time lag to reach maximum $\tilde
T_{a^{\mathrm{t}}_1\rightarrow a^{\mathrm{p}}_1}$ and $\tilde
T_{a^{\mathrm{p}}_1\rightarrow a^{\mathrm{t}}_1}$ is $\Delta
tU_b/D\approx2$. This delay can be interpreted as the elapsed time of
the fluid in porous media to the influence the free flow, or vice
versa. For increasing $l$, the time delay for maximum transfer
entropy increases linearly with $l$ along the upper ridge
(solid lines) with a slope of $l/\Delta t\approx 0.4U_b$. This linear increase of the time delay can be explained by the
convection of the turbulent structures from the position of
source signal to the target signal at a mean speed of $0.4U_b$.

For $l\ge1$, a second ridge appears in figure \ref{fig:Tij}
(dashed lines). For $\tilde T_{a^{\mathrm{p}}_1\rightarrow
  a^{\mathrm{t}}_1}$, the information transfer along this second ridge
is stronger than along the first ridge discussed above, suggesting that
the `bottom-up' coupling is stronger between neighbouring pore
units. The emergence of the second ridge could be attributed to the linkage of neighbouring pore units. As shown in figure \ref{fig:inst_u}, a local upwelling/downwelling events are usually accompanied by a downwelling/upwelling event at nearby pore units, owing to the continuity constraint. This leads to a general phase shift between the fluctuations in neighbouring pore units and also a strong coupling effect between them. Such a coupling of neighbouring pore units make it possible for the the upstream sub-surface fluctuations to affect downstream free flow indirectly, which is reflected on the second ridge in figure \ref{fig:Tij}(a2,b2). 

Despite the similarities observed for different porosities, there are
striking differences between the two cases shown in figure
\ref{fig:Tij}. The magnitude of $\tilde T$ as well as its time and
spatial extent are larger for case C08 (i.e., larger porosity) than
for case C05.  The results might be explained by noting that, first,
the gap of cylinders in C05 is covered by a strong vortex, which
blocks the free flow from entering the porous medium region directly
(as shown in figure \ref{fig:POD_A}a). On the other thand, the gap is
free from any obstructing vortex for C08 (figure \ref{fig:POD_A}b),
which facilitates the exchange of fluid. The larger time and spatial
extent for high porosity cases can be attributed to the large scale KH
eddies originated from the inflection of mean profile, which is
further studied in the next section. The behavior of cases C06 and C07, omitted for brevity, is situated between C05 and C08.

\subsection{The dependence of causality on time scale}

\begin{figure}
\centering
\includegraphics[width=1\textwidth]{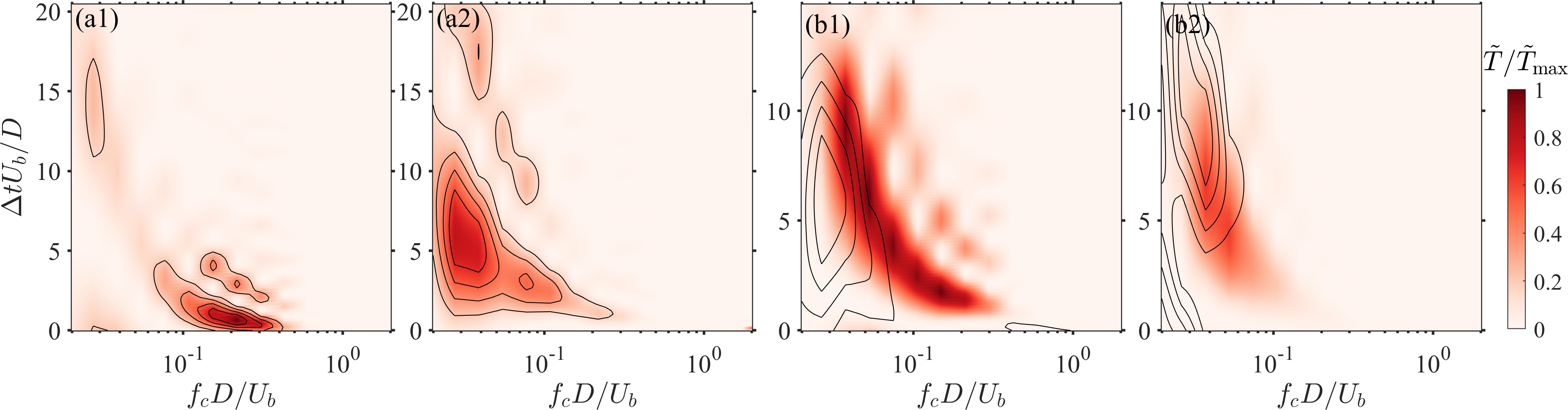}
\caption{Dependence of transfer entropy on time scale for (a1,a2) case
  C05 and (b1,b2) case C08. (a1,b1) colored contour: $\tilde T_{\check
    a^{\mathrm{t}}_1\rightarrow a^{\mathrm{p}}_1}$, isolines: $\tilde
  T_{\check u^{\mathrm{t}}\rightarrow M^{\mathrm{p}}}$ ; (a2,b2)
  colored contour: $\tilde T_{a^{\mathrm{p}}_1\rightarrow \check
    a^{\mathrm{t}}_1}$, isolines: $\tilde
  T_{M^{\mathrm{p}}\rightarrow\check u^{\mathrm{t}} }$. The levels of
  isolines are from 0.2 to 0.8 with a step of 0.2. }
\label{fig:Tij_scale}
\end{figure}

The fluctuations in the turbulent free flow consist of multiple time
scales as illustrated by the POD time coefficients shown in figure
\ref{fig:POD_A}(d). In this section, we investigate which of these
scales have largest impact on the interactions between turbulent and
sub-surface flow. To address this point, a band-pass filter is
applied to $a^{\mathrm{t}}_1$ to select modes in a certain frequency
range,
\begin{equation}
    \check a^\mathrm{t}_1(t)=\int_{f_c-0.5B}^{f_c+0.5B} \hat
    a^\mathrm{t}_1(f) e^{2\pi \mathrm{i} ft}\mathrm{d}f +
    \int_{-f_c-0.5B}^{-f_c+0.5B} \hat a^\mathrm{t}_1(f) e^{2\pi
      \mathrm{i} ft}\mathrm{d}f
\end{equation}
where $(\check \cdot)$ indicates a band-pass filtered signal; $(\hat
\cdot)$ denotes the Fourier transformation in time; $f_c$ is the
central frequency of the filter band and $B$ is the bandwidth. The
bandwidth is set to $B D/U_b=0.1$, which provides a balance between
localised frequencies and statistical convergence of the results. The
transfer entropy as defined in (\ref{eq:nte}) is then evaluated
between $\check a^{\mathrm t}_1$ and $a^{\mathrm{p}}_1$. The results
(colored contour in figure \ref{fig:Tij_scale}) reveal fundamental
differences between top-down and bottom-up interactions as a function
of porosity.

For case C05, the top-down information transfer $\tilde T_{\check
  a^{\mathrm{t}}_1\rightarrow a^{\mathrm{p}}_1}$ is mainly carried out
by modes within the range $f_cD/U_b\approx0.1$--$0.3$ (figure
\ref{fig:Tij_scale}a1), which translates into a streamwise wavelength
of $\lambda_x \approx U_b u_\tau^\mathrm{p}/(f_c\nu) \approx
300$--$1000$ in wall units using Taylor's hypothesis. This is close to
the length scale of near-wall streamwise vortices and streaks,
suggesting that near-wall structures have a considerable impact on the
momentum transport inside porous medium. On the other hand, bottom-up
transfer $\tilde T_{a^{\mathrm{p}}_1\rightarrow \check
  a^{\mathrm{t}}_1}$ mostly resides on the low-frequency modes
$f_cD/U_b\approx0.02$--$0.05$ (figure \ref{fig:Tij_scale}a2), which
suggests that the main effect of the sub-surface flow on the upper
turbulent boundary layer is the disruption of large-scale structures
by upwelling/downwelling events. This is consistent with figure
\ref{fig:Tij} (a1,a2) where bottom-up information transfer persists
for longer distances than top-down events.


The scale dependence in case C08 is notably different from case
C05. In the former, low frequency modes ($f_cD/U_b<0.1$) dominate both
 top-down and bottom-up information transfer. The characteristic
frequency of shear instability can be estimated for case C08 by the
relation suggested by \citet{ghisalberti2002mixing}
as $f_{\mathrm{KH}}D/U_b\approx 0.04$, which is consistent with the
active large-scale modes in figure
\ref{fig:Tij_scale}(b1,b2). Moreover, the time sequence of
$a^\mathrm{t}_1$ and $a^\mathrm{p}_1$ in figure \ref{fig:POD_A}(d)
also hints at low frequency fluctuations close to
$f_{\mathrm{KH}}$. These observations suggest that the near wall
turbulence in case C08 is dominated by large-scale shear instability
modes, which give rise to strong surface-subsurface interactions. The
presence of KH eddies in C08 overwhelms the role of near-wall
attached eddies and demonstrates that the change of wall permeability
can fundamentally altered the interaction scheme. The cases with medium
porosity C06 and C07 exhibit a behavior closer to C08 in terms of scale
dependency, and they are not shown here for the sake of brevity.

The POD time coefficients $a^{\mathrm{t}}_1$ and $a^{\mathrm{p}}_1$
used in the analysis above may lack information on very high-frequency
fluctuations. For completeness, we assess the influence on the results
of using signals with additional temporal spectral content.  For the
free flow region, we define the instantaneous turbulent fluctuation
$u^{\mathrm t}=u'(x_c,y_{\mathrm{peak}},z_c,t)$ from the peak of
$\langle \overline{q}\rangle(y)$ ( $y_{\mathrm{peak}}^+\approx15$)
above the gap center $(x_c,y_c,z_c)$. The instantaneous vertical mass
flux across cylinder gaps $M^{\mathrm{p}}(t)=\int_{\mathrm{gap}}
v'(x,t) \mathrm{d}x$ is used to represent the temporal fluctuation in
porous region caused by upwelling/downwelling events. The transfer
entropy between band-pass filtered $\check u^{\mathrm{t}}$ and
$M^{\mathrm{p}}$, $T_{\check u^{\mathrm{t}}\rightarrow M^{\mathrm{p}}
}$ and $T_{M^{\mathrm{p}}\rightarrow\check u^{\mathrm{t}} }$, is then
calculated following the same procedure as above, and the results are
superimposed in figure \ref{fig:Tij_scale} as solid isolines. For case
C05, the result for the two sets of signals is essentially
identical. For case C08, the patterns of $T_{\check
  u^{\mathrm{t}}\rightarrow M^{\mathrm{p}} }$ and $T_{M^{\mathrm{p}}\rightarrow\check u^{\mathrm{t}} }$ shift even more
toward low frequencies ($f_cD/U_b<0.1$). Hence, both contours
emphasise the role of large scale KH eddies, and the conclusions drawn
are consistent with the analysis using POD coefficients.

\section{Concluding remarks}\label{sec:conclusion}

We have investigated the causal interactions between the turbulent
flow over a porous media and the sub-surface flow using information
theoretic tools.  The data was obtained by interface-resolved direct
numerical simulations and transfer entropy was employed to quantify
the causality between time signals representative of the flow
structure. A collection of time series of POD coefficients and
instantaneous velocity/momentum flux signals have been used to
characterise the dynamics of energy-containing eddies in the free flow
and the subsurface flow region. We have focused on two effects: i) the
bottom-up and top-down directionality of the interactions across the
surface-subsurface interface and ii) the impact of the media porosity
on the nature of these interactions.

Our results show that the porosity of the porous medium has a profound
impact on the intensity, time scale and streamwise extent of
surface-subsurface interactions. For values of porosity equal to 0.5,
there is a clear asymmetry between top-down and bottom-up
interactions. The former are dominated by the influence of near wall
attached eddies (e.g. streamwise vortices and streaks) on the surface
flow, whereas the latter are mostly the disruption of free flow
large-scale structures by upwelling/downwelling events.  As the
porosity increases, bottom-up interactions remain unchanged; however,
the flow structures responsible for top-down interactions change from
near-wall attached eddies to large scale shear-instability eddies,
leading to an increase in temporal and spatial extent of causal
interactions.  This suggests that the perturbation induced by the
vertical momentum flux across the interface is an important source of
shear instabilities in flows over porous beds.

\begin{acknowledgements}
The study has been financially supported by the Deutsche
Forschungsgemeinschaft (German Research Foundation) project SFB1313
(project No. 327154368). The authors report no conflict of interest.
\end{acknowledgements}
\bibliographystyle{jfm}
\bibliography{references}
\end{document}